\documentclass[twocolumn,10pt]{article}
\usepackage[utf8]{inputenc}
\usepackage{amsmath,amssymb,amsfonts}
\usepackage{graphicx}
\usepackage{physics}
\usepackage{xcolor}
\usepackage{makecell}
\usepackage{siunitx}
\usepackage{booktabs}
\usepackage{tikz}
\usepackage{pgfplots}
\usepackage{subcaption}
\usepackage{tabularx}
\usepackage{algorithm}
\usepackage{algpseudocode}
\usepackage{setspace}
\usepackage{url}
\usepackage{array}
\usepackage[margin=0.75in]{geometry}
\usepackage{pgfplotstable}
\usetikzlibrary{3d,positioning,angles,quotes}
\pgfplotsset{compat=1.17}

\newcolumntype{C}{>{\centering\arraybackslash}X}

\title{Maximum Likelihood Estimation Yields Accurate Line-of-Response Assignment for Positron + Prompt Gamma Ray Events in Multiplexed PET (mPET)}

\author{%
Sarah J. Zou\textsuperscript{1--3}, Garry Chinn\textsuperscript{2,3}, Muhammad Nasir Ullah\textsuperscript{2,3}, Craig S. Levin\textsuperscript{1--5*}\\[0.5em]
\textsuperscript{1}Department of Electrical Engineering, Stanford University\\
\textsuperscript{2}Department of Radiology, Stanford University\\
\textsuperscript{3}Molecular Imaging Program at Stanford (MIPS)\\
\textsuperscript{4}Department of Bioengineering, Stanford University\\
\textsuperscript{5}Department of Physics, Stanford University\\[0.5em]
\textsuperscript{*}Corresponding author: Craig S. Levin, cslevin@stanford.edu
}

\date{}

\begin{document}

\makeatletter
\twocolumn[
\begin{@twocolumnfalse}
\maketitle

\begin{abstract}
For accurate disease characterization using positron emission tomography (PET), it is desirable to image multiple radiotracers in a single scan. Conventional PET methods cannot do this due to the indistinguishable annihilation photons produced by different radiotracers. One approach is to label one radiotracer with a positron+prompt-gamma ($\beta^+\!\!-\!\!\gamma$) isotope producing triple coincidences, and another with a pure positron-emitting ($\beta^+$) isotope producing double coincidences. However, $\beta^+\!\!-\!\!\gamma$ emitters present challenges in correctly identifying the two annihilation photons, or equivalently, assigning the correct line-of-response (LOR) to triple-photon coincidence events. Here, we propose a maximum likelihood estimation (MLE) framework leveraging spatial, timing, and energy information to determine the most probable LOR. Simulation studies validated the method: simulations showed over 96\% and 94\% accuracy for LOR assignment of $\beta^+\!\!-\!\!\gamma$ emitters $^{22}$Na and $^{124}$I point sources, respectively. Furthermore, simulated phantom imaging of $^{22}$Na or $^{124}$I distributions alongside a $\beta^+$ emitter demonstrated that MLE LOR assignment achieved comparable image quality---measured by contrast recovery coefficient (CRC) and cross-talk ratio (XR)---to benchmark methods, where the prompt gamma was identified using an energy threshold ($\geq 650$ keV) for $^{22}$Na and as the highest-energy photon for $^{124}$I.
\end{abstract}

\vspace{1em}
\end{@twocolumnfalse}
]
\makeatother

\section{Introduction}
Positron emission tomography (PET) is a non-invasive medical imaging modality that visualizes and quantifies molecular pathways of disease in living 
subjects \cite{muehllehner_positron_2006}. By tracking a radioactively labeled contrast agent (radiotracer), PET reveals molecular and cellular processes, such as glucose consumption, receptor 
status, and DNA proliferation, generating quantitative images of the tracer's distribution in specific regions \cite{trotter_positron_2023}. However, conventional PET systems are limited to 
imaging a single radiotracer per session owing to their reliance on detecting pairs of 511 keV positron-electron annihilation photons emitted from the radiotracer \cite{black_rapid_2009}. 
These photons are detected along lines of response (LOR) formed by pairs of opposing detector elements in the PET ring. The 3D event patterns collected across all LORs are then used to reconstruct the tracer's biodistribution \cite{wernick2004emission}. \par

Numerous radionuclides exhibit an often-overlooked property: they emit an additional gamma ray in cascade with the positron, classifying them as positron+prompt-gamma ($\beta^+\!\!-\!\!\gamma$) emitters. Examples include \(^{124}\mathrm{I}\), \(^{22}\mathrm{Na}\), \(^{44}\mathrm{Sc}\), and \(^{60}\mathrm{Cu}\); for tables of $\beta^+$ and $\beta^+\!\!-\!\!\gamma$ emitters, see \cite{pratt_simultaneous_2023, tashima_three-gamma_2024}. $\beta^+\!\!-\!\!\gamma$ emitters used in this work are $^{22}$Na and $^{124}$I, which emit prompt gamma rays of 1275~keV and 602~keV, respectively, in cascade
with positron emission. Traditionally, these prompt gammas produce random coincidences, prompting the use of narrow energy windows and advanced signal processing methods to avoid contamination \cite{pentlow_quantitative_1996}. In principle, PET systems are capable of detecting the prompt-gamma from $\beta^+\!\!-\!\!\gamma$  emitters which can be used to differentiate them from pure positron ($\beta^+$) emitters. This capability enables simultaneous imaging of multiple tracers in a single PET session, a technique referred to as multiplexed PET (mPET) \cite{andreyev_dual-isotope_2011, gonzalez_methods_2011}. 
mPET facilitates a comprehensive assessment of multiple disease biomarkers, which can be leveraged for improved diagnostic accuracy and more effective treatment. For instance, in oncology, different tracers can characterize tumors based on glucose metabolism, hypoxia, blood flow, and cellular proliferation, aiding in the determination of optimal treatment strategies \cite{kadrmas_methodology_2013}. While these biomarkers can be assessed through sequential scans, simultaneous dual-tracer imaging eliminates the logistical challenges of multiple scans and enables evaluation of all biomarkers at a single time point, avoiding potential temporal drift in biomarker expression \cite{soultanidis_multiplexed_2025}.

In this study, we present a maximum likelihood estimation framework for the accurate assignment of the correct line of response (MLE LOR) in the context of triple coincidences involving $\beta^+\!\!-\!\!\gamma$ emitters. Each triple coincidence event yields three potential LORs, making precise assignment crucial for optimal image reconstruction of the $\beta^+\!\!-\!\!\gamma$ emitter. 

\subsection{Previous Work}
Triple coincidences, in which three photons are detected within a coincidence time window, can arise from $\beta^+\!\!-\!\!\gamma$ isotope decay or from random events involving both pure $\beta^+$ and $\beta^+\!\!-\!\!\gamma$ emitters. Correctly assigning a LOR for a triple coincidence is important for various applications; here, we focus specifically on triple coincidences from $\beta^+\!\!-\!\!\gamma$ emitters in the context of mPET. Further applications of three-gamma imaging are discussed in this review \cite{tashima_three-gamma_2024}.  

Dual-isotope PET with  $\beta^+\!\!-\!\!\gamma$ emitters have been explored through simulation studies where reconstruction methods and sensitivity estimations have been reported \cite{andreyev_dual-isotope_2011}. 
In this work, Andreyev \textit{et. al.} identified the prompt-gamma based on an energy threshold of $\geq 650$ keV and used an energy window of 460-560 keV for annihilation photons. Another similar energy-based approach was done in the context of imaging three isotopes simultaneously where the max energy photon in a triple coincidence was assumed to be the prompt-gamma \cite{zou_triplexed_2025_1}. Lage \textit{et al.} \cite{lage2015recovery} uses double coincidence data to weight candidate LORs in a triple coincidence in the context of imaging a $\beta^+$ emitter and boasting sensitivity by using the triple coincidences that would normally be thrown away. This method is not suitable for mPET involving both pure $\beta^+$ and $\beta^+\!\!-\!\!\gamma$ emitters since both emitters contribute to the double coincidence set and will likely result in $\beta^+$ signal in the triple coincidence image.

The concept of measuring double and triple coincidences was explored in a three detector configuration experiment with $^{68}$Ge and $^{22}$Na point sources in 2011 \cite{miyaoka_dual-radioisotope_2011}. In this study, the authors used 2 collinear detectors to detect annihilation photons and had a prompt-gamma only detector (lower energy threshold of 850 keV) to differentiate between double and triple coincidences. A similar approach for mice imaging of $^{22}$Na and $^{18}$F was done in \cite{fukuchi_positron_2017}.

In 2019, Moore \textit{et al.}, simultaneously imaged $^{18}$F and $^{124}$I in NEMA-NU4 phantom in a preclinical scanner (Molecubes $\beta$-CUBE $\mu$PET) \cite{moore2019simultaneous} where triple coincidences were recorded as sequential double coincidences with a shared detector and all possible LORs were reconstructed. A similar approach was also done in \cite{pratt_simultaneous_2023, zou_quantitative_2025} with the Siemens Inveon preclinical imaging system.

Our group at Stanford University (Molecular Imaging Instrumentation  (MIIL)) previously developed a statistical method for identifying the LOR for $\beta^+ - \gamma$ decay events \cite{Elyssa_MLE}. This approach uses both energy and timing information to determine the most likely LOR during a triple coincidence event.
    
\subsection{Original Contribution}
This work builds upon the previous statistical framework presented in \cite{Elyssa_MLE} by refining the time probability to incorporate spatial resolution. In this work, we perform a comprehensive evaluation of our algorithm over a range of time and energy resolutions for both $^{22}$Na and $^{124}$I point sources in simulation. Additionally, we demonstrate performance in simulated mPET acquisition of a $\beta^+$ emitter with $^{22}$Na or $^{124}$I. While ${}^{22}$Na has limited clinical relevance, it is used here as a representative positron–prompt-gamma emitter with a high prompt-gamma energy (1275 keV), complementing ${}^{124}$I (602 keV). Together, these isotopes span a range of prompt-gamma energies relevant to multiplexed PET and enable evaluation of the proposed method across different emission characteristics.

\section{Algorithm Overview}
This section summarizes the process of identifying the most probable LOR from a detected true triple coincidence event. Each event provides three key pieces of information: the positions of the detectors $\mathbf{p} = (\mathbf{p_1}, \mathbf{p_2}, \mathbf{p_3})$, the energies detected by each detector $\mathbf{e} = (e_1, e_2, e_3)$, and single photon time of arrival information $\mathbf{t} =(t_1, t_2, t_3)$.

From these inputs, there are three possible LORs to be assigned to this event by connecting pairs of detector positions: $L_1 = (\mathbf{p_1}, \mathbf{p_2}), L_2 = (\mathbf{p_2}, \mathbf{p_3}), L_3 = (\mathbf{p_1}, \mathbf{p_3})$.The first step of the algorithm is to filter out LORs that do not intersect the imaging field of view (FOV).

For the remaining LORs, we calculate their likelihood of representing the actual event using two criteria. First, we evaluate the energy probability (Section~\ref{sec:energy_prob}), which measures how well the detected energies along the LOR match the expected energy distribution. Second, we assess the spatial and timing probability (Section~\ref{sec:time_space_prob}) by analyzing the time differences between signals in the detectors, location of detection, and scanner spatial resolution. These two probabilities are combined to identify the LOR that  most likely corresponds to the annihilation event.

We aim to select the LOR with the highest posterior probability $P(L_i | \mathbf{p}, \mathbf{t},\mathbf{e})$, which represents the likelihood that a given LOR $L_i$ corresponds to the 511 keV photon pair, given the detector positions, times, and energies. By applying Bayes' rule, we express this probability as:
\begin{equation}
\begin{aligned}
    P(& L_i  \,|\, \mathbf{p},\mathbf{t},\mathbf{e}) \\
    &= P(\mathbf{p}, \mathbf{t},\mathbf{e} \,|\, L_i) \frac{P(L_i)}{P(\mathbf{p},\mathbf{t},\mathbf{e})}
\end{aligned}
\end{equation}

Where: 
\begin{itemize}
    \item \( P(\mathbf{p}, \mathbf{t}, \mathbf{e} | L_i) \) is the likelihood, representing the probability of observing the positions, times, and energies given the correct LOR is \( L_i \).
    \item \( P(L_i) \) is the prior probability of LOR \( L_i \), we assume equal probability of each LOR in the FOV.
    \item \( P(\mathbf{p}, \mathbf{t}, \mathbf{e}) \) is the evidence or normalization constant that has no bearing on if one LOR is more likely than the other.
\end{itemize}

With this assumption, the posterior probability \( P(L_i | \mathbf{p}, \mathbf{t}, \mathbf{e}) \) becomes directly proportional to the likelihood \( P(\mathbf{p}, \mathbf{t}, \mathbf{e} | L_i) \). 
Assuming energy is conditionally independent from time and position, the probability can be simplified as:

\begin{equation}
\begin{aligned}
    P(& \mathbf{p},\mathbf{t},\mathbf{e} \,|\, L_i) \\
    &= P(\mathbf{e} \,|\, L_i) \, P(\mathbf{p}, \mathbf{t} \,|\, L_i)
\end{aligned}
\end{equation}

\begin{figure}[htbp]
  \centering
  \includegraphics[width=\columnwidth]{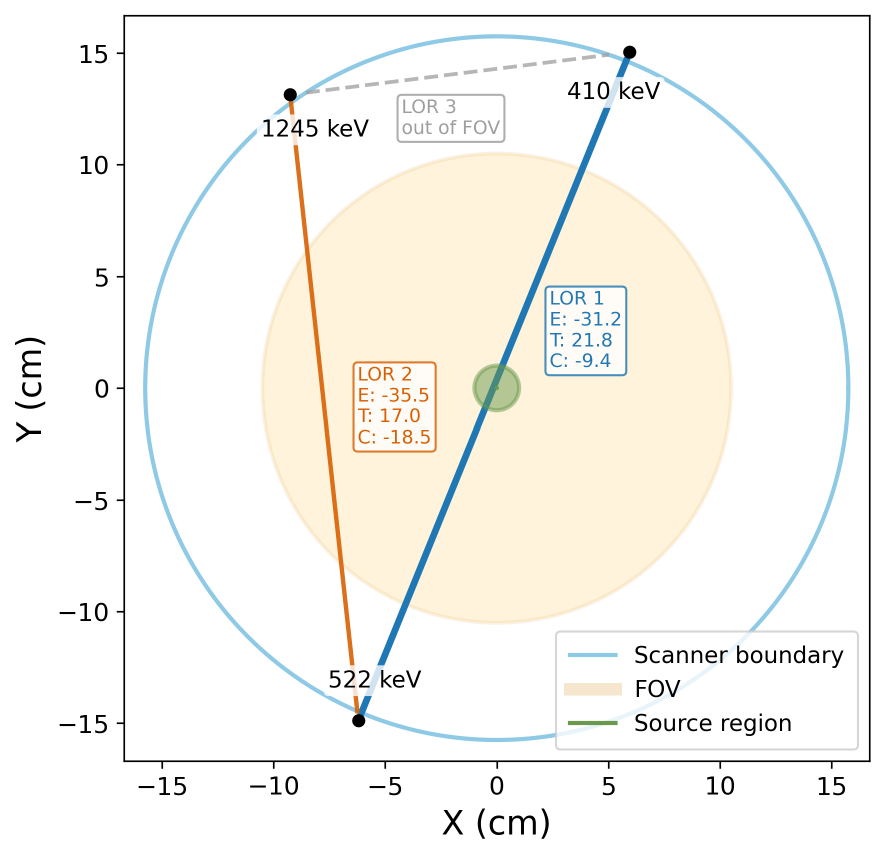}
    \caption{\raggedright
    Visualization of three candidate lines of response (LORs) from a single triple coincidence event in a 2D projection of the PET scanner. The scanner boundary (blue circle), imaging field of view (FOV, shaded region), and source region (green circle) are shown. A $^{22}$Na point source is located near the center. Three possible LORs are illustrated: LOR 1 (blue), LOR 2 (orange), and LOR 3 (gray dashed), with LOR 3 excluded for falling outside the FOV. Detected photon energies (in keV) are labeled at each interaction point. For each valid LOR, the log-probabilities of energy (E), timing (T), and their combination (C) are displayed. The true LOR (LOR 1) yields the highest combined log-probability, demonstrating the principle of the maximum likelihood estimation (MLE) framework for LOR assignment.
    }
  \label{fig:lor_example}
\end{figure}

The overview of the approach is presented in Algorithm \ref{MLE_algo}. An example of energy based probability and timing probability of each possible LOR is shown in Fig.~\ref{fig:lor_example}.

\begin{algorithm} 
\caption{Triple Coincidence LOR} \label{MLE_algo}
\begin{algorithmic}[1]
    \Procedure{find\_LOR}{triple\_event} \Comment{triple\_event is a list of length three of dictionaries of each detected hit's time, energy, and coordinates}
        \State $LORS \gets \text{make\_possible\_LORS(triple\_event)}$
        \State $LORS \gets \text{exclude\_FOV(LORS)}$
  
        \If{$\text{len}(LORS) == 0$}
            \State \textbf{return} None
        \Else
            \State lor\_probability $\gets$ \text{energy\_probabilities(LORS, energies)} + \text{time\_and\_spatial\_probabilities(LORS, positions,  times)}
            \State max\_ind $\gets$ \text{max\_arg(lor\_probability)}
            \State \textbf{return} $LORS[\text{max\_ind}]$
        \EndIf
    \EndProcedure
\end{algorithmic}
\end{algorithm}

\subsection{Energy Probability} \label{sec:energy_prob}
The energy distributions were modeled as normal distributions centered at the annihilation energy (511 keV), denoted $P_\text{ann}$, and the prompt-gamma energies (1275 keV for $^{22}$Na and 602 keV for $^{124}$I), denoted $P_\gamma$. The standard deviation $\sigma$ of each distribution was calculated as $\sigma = \text{FWHM} / 2.355 = (\text{energy resolution (\%)}) \times E / 2.355$, where $E$ is the distribution’s center.  Owing to the prompt-gamma energy of 22Na (1275 keV) being far apart from 511 keV annihilation energy and to avoid near zero probabilities for $P_\gamma$, $P_\gamma(e \leq 750) = P_\gamma(e = 750)$ was assumed for $^{22}$Na energy calculations. 

The energy likelihood for an LOR candidate can be estimated as:  
\begin{equation}
    P(\mathbf{e} \,|\, L_i)
     = P_\text{ann}(e_i) P_\text{ann}(e_{i+1}) P_\gamma(e_{i+2})
\end{equation}

\subsection{Spatial and Timing Probability} \label{sec:time_space_prob}
For spatial and timing probability (hereafter referred to as the timing probability), we find the probability that we observe the recorded position and single photon arrival times $\mathbf{p}, \mathbf{t}$ given that $L_i$ is the correct line of response or $P(\mathbf{p}, \mathbf{t} | L_i)$.  The probabilities were calculated as a function of latent variables representing the true annihilation position, time, and photon travel vectors. 
Given the hypothesis that $\mathbf{p}_1, \mathbf{p}_2$ are the position of the detected annihilation photons, we estimate the latent variables $\mathbf{x}, t_0, \mathbf{v}_1, \mathbf{v}_2, \mathbf{v}_3$ where $\mathbf{x}$ is the coordinates of the annihilation event, $t_0$ is the time of the annihilation event, and $\mathbf{v}_1, \mathbf{v}_2, \mathbf{v}_3$ are the vectors that show the distance and direction of travel of each photon from decay location to detector. We assume oppositely-directed annihilation photons, or
$\hat{\mathbf{v}}_1 \cdot \hat{\mathbf{v}}_2 = -1$. 

The LOR $L_i$ observed from an annihilation event is fully determined by these latent variables $(\mathbf{x}, t_0, \mathbf{v}_{1-3})$. Owing to this dependence, $P(\mathbf{p}, \mathbf{t} | L_i) = P(\mathbf{p}, \mathbf{t} | \mathbf{x}, t_0, \mathbf{v}_1, \mathbf{v}_2, \mathbf{v}_3)$.

\subsubsection{Latent Variable Estimation}

The estimation of latent variables can be derived by :
\begin{equation}\label{eq:x_int}
   \mathbf{\hat{x}} = \frac{\mathbf{p}_1 + \mathbf{p}_2}{2} + \frac{c \Delta t}{2} \frac{(\mathbf{p}_{2} -\mathbf{p}_{1})}{\norm{\mathbf{p}_{2} -\mathbf{p}_{1}}}
\end{equation}
where 
\begin{equation}
\Delta t = t_{1} - t_{2}
\end{equation}
In Eq. \ref{eq:x_int}, we assume annihilation photon collinearity and use time of flight to estimate location of the annihilation event along the LOR. The following variables are initialized as ($c_\text{mm}$ is the speed of light expressed in millimeters per second):
\begin{align}
\mathbf{\hat{v}}_1 &= \mathbf{p}_1 - \mathbf{\hat{x}} \\
\mathbf{\hat{v}}_2 &= \mathbf{p}_2 - \mathbf{\hat{x}} \\
\mathbf{\hat{v}}_3 &= \mathbf{p}_3 - \mathbf{\hat{x}} \\
\hat{t}_0 &= t_2 - \frac{\|\mathbf{\hat{v}}_2\|}{c_{\text{mm}}}
\end{align}

\subsubsection{Probability Calculation}
Assuming conditional independence between the latent variables, we can then decouple the time spatial probability accordingly:
\begin{align} \label{eq: prob_decouple}
    P(&\mathbf{p}_1, \mathbf{p}_2, \mathbf{p}_3, t_1, t_2, t_3 \mid \mathbf{x}, t_0, \mathbf{v}_1, \mathbf{v}_2, \mathbf{v}_3) \nonumber \\
    &= P(\mathbf{p}_1 \mid \mathbf{x}, \mathbf{v}_1) \cdot P(\mathbf{p}_2 \mid \mathbf{x}, \mathbf{v}_2) \cdot P(\mathbf{p}_3 \mid \mathbf{x}, \mathbf{v}_3) \nonumber \\
    &\quad \cdot P(t_1 \mid t_0, \mathbf{v}_1) \cdot P(t_2 \mid t_0, \mathbf{v}_2) \cdot P(t_3 \mid t_0, \mathbf{v}_3)
\end{align}

We then assume Gaussian probability:
For $i = 1, 2, 3$:

\begin{equation} \label{eq:pos_gaussian}
    P(\mathbf{p}_i \mid \mathbf{x}, \mathbf{v}_i)\sim \mathcal{N} (\mathbf{x} + \mathbf{v}_i, \Sigma)
\end{equation}
Where $\mathcal{N}$ represents a multivariate Gaussian with covariance matrix of $\Sigma = \begin{bmatrix}
    \sigma_{x} & 0 & 0 \\
    0 & \sigma_{y} & 0 \\
    0 & 0 & \sigma_{z}
\end{bmatrix}$ where $\sigma_x, \sigma_y, \sigma_z$ are the spatial resolution on the respective axis. 
\begin{equation}
    P(t_i \mid t_0, \mathbf{v}_i) \sim \mathcal{N} (t_0 + \frac{\norm{\mathbf{v}_i}}{c_\text{mm}}, \sigma_t)
\end{equation}

Where $\mathcal{N}$ is one-dimensional, $\sigma_t$ is the standard deviation of the singles time resolution (corresponding to simulated FWHM values of 100--1000 ps), and $c_\text{mm}$ is the speed of light in millimeters.

\section{Methods}
 The proposed algorithm, was validated and its performance assessed through Monte Carlo simulations (GATE).

\subsection{GATE Simulations}
The Monte Carlo simulations were performed using GATE version 8.2 \cite{jan_gate_2004}. The PET scanner simulated for this work is our lab built MR-compatible PETcoil2 with inner diameter of 315 mm and axial FOV of 160 mm \cite{dong_petcoil_2024}. The FOV region is defined to be a cylinder of 10.5 cm radius and 16 cm length. The system detectors comprise LYSO crystal elements, each with dimensions of 3.2 × 3.2 × 20 mm. There are 16 detectors in a ring with 768 crystals per detector module. We collected singles data with upper energy threshold of 1500 keV and a lower energy threshold of 460 keV. The adder used the \texttt{TakeEnergyCentroid} policy, which averages multiple hits and records an energy-weighted interaction position. From the singles data, we collected the position $(x, y, z)$, energy, time, and eventID. $^{22}$Na was simulated using an ion source type. $^{18}$F was simulated as a back-to-back source of 511 keV, which provides a reasonable and efficient approximation since our goal is to evaluate the method’s effectiveness in imaging the triple emitter in the presence of a $\beta^+$ emitter.

\subsubsection{${}^{22}$Na and ${}^{124}$I Point Sources}
We simulated a 1~mm radius point source with 16,818~Bq activity centered within a 10~mm radius spherical water phantom using GATE. The source was placed at three positions along the x-axis: (0, 0, 0), (50, 0, 0), and (100, 0, 0)~mm. These simulations were repeated with the source shifted axially to $z = 40$~mm (one-quarter of the axial FOV). For ${}^{22}$Na, a 10-second acquisition was acquired for each position. The same simulation setup was repeated for ${}^{124}$I using the \texttt{fastI124} model, with a 100-second acquisition to account for its lower positron branching ratio (23\%) and 51\% prompt-gamma emission rate, compared to ${}^{22}$Na's 90\% positron branching ratio and 100\% prompt-gamma emission \cite{tashima_three-gamma_2024}.

\subsubsection{Heterogeneity Phantom} \label{sec:hetero_phantom}
\begin{figure}[htbp]
    \centering
    \includegraphics[width=\columnwidth]{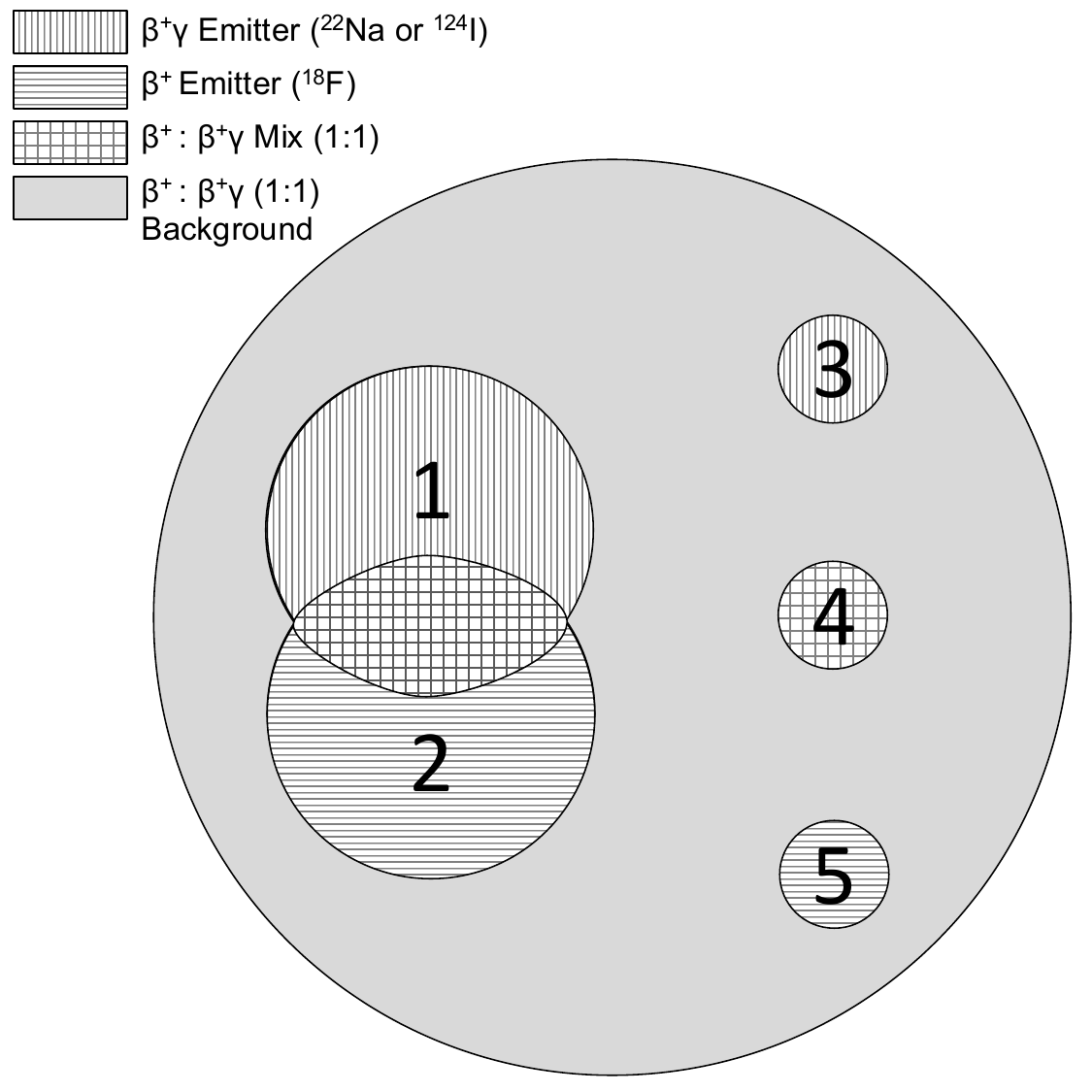}
    \caption{The heterogeneous radionuclide phantom used in GATE simulations with ROI labels. The total size of the phantom is a cylinder of radius 10.5 cm and height of 20 cm. The small spheres have diameter of 2 cm and a 4:1 hot-to-warm background activity concentration ratio and large spheres have diameter of 6 cm and a 3:1 hot-to-warm background activity concentration ratio.}
    \label{fig:6_sphere_layout}
\end{figure}

To demonstrate the algorithm's suitability for mPET—where a patient is injected with two tracers simultaneously, potentially with overlapping spatial distributions—we simulated a phantom configuration with overlapping sources, as shown in Fig.~\ref{fig:6_sphere_layout}. The configuration consists of three small spheres with a radius of 1~cm: two containing a single radionuclide and one containing a 1:1 mixture of both. In addition, two larger, partially overlapping spheres with a radius of 3~cm were included.

All sources were embedded in a warm cylindrical water phantom (radius 10.5 cm, height 20 cm) containing a uniform background of $^{18}$F and $^{22}$Na at equal concentrations. Small spheres (ROIs 3 and 5 in Fig.~\ref{fig:6_sphere_layout}) containing a single radionuclide were simulated at a concentration corresponding to a 4:1 hot-to-background activity concentration ratio. The mixed sphere (ROI 4 in Fig.~\ref{fig:6_sphere_layout}) contained both radionuclides at the same concentration. Large spheres (ROIs 1 and 2 in Fig.~\ref{fig:6_sphere_layout}) were simulated at a concentration corresponding to a 3:1 hot-to-background activity concentration ratio. An additional simulation was performed using the same phantom and activity distribution, but with $^{124}$I (using GATE's \texttt{fastI124} model) instead of $^{22}$Na. 

\subsubsection{Data Processing}
We blurred the collected data to simulate a range of time, energy, and spatial resolutions. 
We applied Gaussian blur of 100 ps - 1ns FWHM on time field, 3 - 4 mm FWHM spatially, and 12 and 18\% FWHM energy at 511 keV. We wanted to cover both TOF ($\leq$ 500 ps singles time resolution (STR) FWHM)  and non-TOF ($>$ 500 ps STR) PET capabilities and encompass energy resolution of both LYSO  (12\% at 511 keV) and BGO (18\% at 511 keV) crystals \cite{spanoudaki_photo-detectors_2010, berg_innovations_2018}. The energy resolution was assumed to be uniform over energy range when in reality energy resolution should improve with higher energies. $4\,\text{mm}$ FWHM represents the chosen effective system resolution for simulation and is a common parameter applied to both isotopes to isolate the comparison of the reconstruction methods. We identified triple coincidences as three single photons with the same eventID. We then used MLE LOR to assign LORs to the triple coincidences.  We then determine accuracy of the selected LOR based on minimum distance between LOR and the point source location. 

For heterogeneity phantom data, we did coincidence grouping with blurred singles timestamps (FWHM = 500 ps) with a coincidence time window of 3 ns. We blur energy by 12\% FWHM and position by 3 FWHM mm in all directions. These blurring parameters were chosen to be representative of the general capabilities of a TOF PET system. The coincidence grouped data was divided into double coincidence and triple coincidence datasets. To filter out random triple coincidence events, we required the time log probability (Sec. \ref{sec:time_space_prob}) to be above 19 (see Appendix \ref{sec:time_threshold} for how this was determined). We do not need to use time threshold for point source since we use eventID and avoid random coincidences. The triple coincidence dataset was processed with our proposed method (MLE LOR) and benchmark methods to yield listmode data of triple coincidence data.

\subsection{Accuracy Evaluation} \label{sec:accuracy_calc}
For each time, energy, and spatial resolution, ten different random seeds were run and the accuracy was recorded for each run and then mean and standard deviation were calculated from these runs.  

Accuracy is calculated as the number of correctly assigned LORs divided by the number of triple-coincidence events with at least one feasible LOR. A LOR is considered correct if it passes within the point source region. An event is included in the denominator only if at least one of its three candidate LORs intersects the point source region.

We calculate accuracy based on the energy probability only (Section~\ref{sec:energy_prob}), the time probability only (Section~\ref{sec:time_space_prob}), and the combined (time \& energy) probability, in order to understand how both components of information affect the accuracy of the proposed algorithm.

We also compare our proposed algorithm to benchmark methods, which are described as follows. For $^{22}$Na, we applied the triple coincidence LOR assignment approach from \cite{andreyev_dual-isotope_2011}, which identifies the photon with energy $\geq$650~keV as the prompt gamma and assigns the two lower-energy photons as the annihilation pair. For $^{124}$I, the two lowest-energy photons in each triple coincidence were assigned as the annihilation pair. 

\subsection{Reconstruction and Image Analysis}
GATE simulation images were normalized with uniform cylinder of ${}^{18}$F. The heterogeneity phantom (Section. \ref{sec:hetero_phantom}) triple coincidence listmodes were reconstructed using ordered-subsets expectation maximization (OSEM) via graphics processing units (GPU) \cite{cui2011fully} with 40 iterations and 70 subsets using time-of-flight TOF imaging. 

To evaluate reconstruction accuracy, we use contrast recovery coefficient (CRC) and cross-talk ratio (XR), which incorporate known activity concentrations in the phantom. Direct voxel-wise comparison to ground truth is not straightforward in PET due to ambiguity in image normalization, and such comparisons can depend strongly on the chosen normalization method. CRC and XR therefore provide a more robust and standardized assessment of reconstruction performance. We evaluated the hot ${}^{22}$Na ROIs (ROIs 1, 3, and 4) using CRC and the ${}^{18}$F-only spheres (ROI 2 and 5) using XR. CRC measures how accurately the reconstructed activity reflects the true activity. For a hot lesion with true activity $A_\text{ROI}$ and background activity $A_\text{BG}$, and measured mean counts $C_\text{ROI}$ and $C_\text{BG}$, it is defined as
\begin{equation}
\mathrm{CRC} = \frac{\left(\dfrac{C_\text{ROI}}{C_\text{BG}} - 1 \right)}{\left( \dfrac{A_\text{ROI}}{A_\text{BG}} - 1 \right)}.
\end{equation}
XR quantifies the cross talk of ${}^{18}$F-only spheres into the triple coincidence image and, for an ${}^{18}$F-only ROI with measured counts $C_\text{F18}$ and background counts $C_\text{BG}$, is defined as
\begin{equation}
\mathrm{XR} = \frac{C_\text{F18} - C_\text{BG}}{C_\text{BG}}.
\end{equation}

In this work, reconstruction focuses exclusively on accurate LOR assignment for triple-coincidence events arising from $\beta^{+}$--$\gamma$ emitters; isotope unmixing accuracy or the degree of suppression of $\beta^{+}$--$\gamma$ double coincidences in the pure-positron image are beyond the scope of this study. Double-coincidence events are treated using standard PET reconstruction pipelines, while the triple-coincidence event LOR assignment method developed here provides improved spatial localization of $\beta^{+}$--$\gamma$ events, which in turn facilitates more accurate downstream isotope-separation methods.

\subsubsection{Bootstrap analysis}
Bootstrap resampling was used to estimate uncertainty in CRC and XR measurements. For each list-mode dataset, 100 bootstrap realizations were generated by sampling coincidence events with replacement from the original list-mode file. Each resampled dataset contained the same number of events as the original acquisition.

Images were reconstructed independently for each bootstrap realization using the four reconstruction approaches evaluated in this work: Benchmark, MLE energy \& time, MLE energy-only, and MLE time-only. CRC and XR metrics were computed for each bootstrap reconstruction. The reported CRC and XR values correspond to the mean across the 100 bootstrap reconstructions, and the standard error of the mean (SEM) was calculated from the bootstrap standard deviation.

In addition, statistical significance between methods was assessed using paired bootstrap confidence intervals. For each bootstrap realization, differences in CRC and XR between methods were computed, and 95\% confidence intervals were obtained from the empirical distribution of these differences. Differences were considered statistically significant if the confidence interval excluded zero.

\section{Results}
\subsection{Simulated $^{22}$Na and $^{124}$I Point Sources}
Simulation of $^{22}$Na point source resulted in 1,464-3,147 triple coincidences for each position (0.87 - 1.88 \% sensitivity). The triple-to-double ratio ranged from 0.10 to 0.14. The number of possible LORs and the number of LORs that meet the benchmark conditions (i.e., at least one photon with energy $\geq 650$ keV and the LOR lies within the FOV) are shown in Table~\ref{tab:na22_possible_benchmark_counts}. For $^{124}$I point source, the number of detected triple coincidences ranged from 3,269 to 6,812 (1.62 - 3.38\% sensitivity). The triple/double ratio ranged from 0.066 to 0.085. The number of possible LORs and LORs that meet the benchmark method (at least one LOR in FOV) criteria are shown in Table~\ref{tab:i124_possible_benchmark_counts}. We saw the same performance over 3 mm FWHM spatial resolution as we did for 4 mm FWHM spatial resolution.

\begin{figure*}[htbp]
    \centering
    \includegraphics[width=2 \columnwidth]{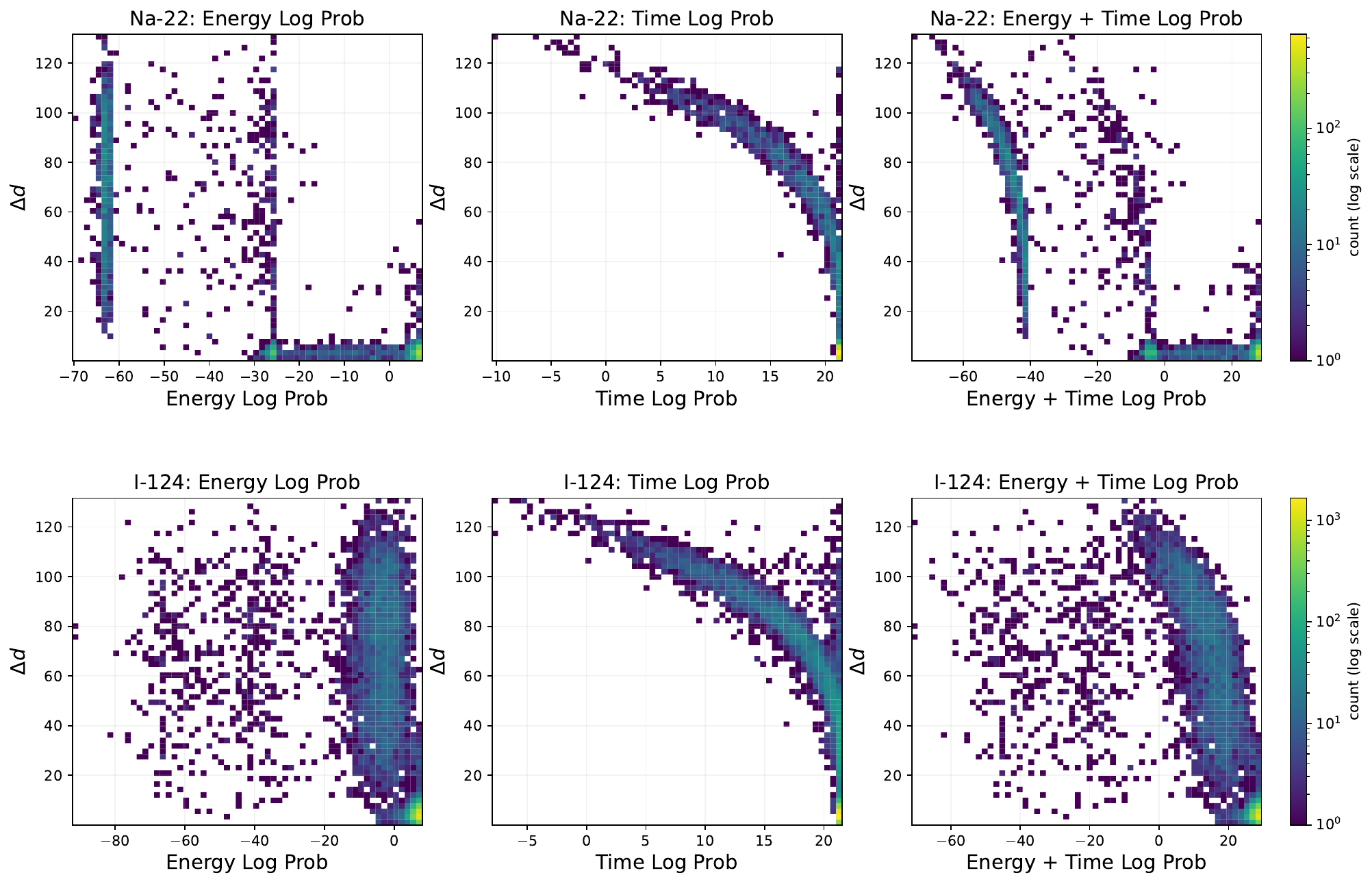}
    \caption{Combined 2D histograms of event log-probability versus $\Delta d$ for Na-22 (top row) and I-124 (bottom row) for point source at (0, 0, 0), 100 ps STR FWHM, 12 \% energy resolution FWHM, and 4 mm spatial FWHM configuration. Columns show Energy Log Prob, Time Log Prob, and Energy + Time Log Prob, respectively. Pixel color indicates event count on a logarithmic scale, with a shared color scale within each isotope row to enable direct comparison across the three probability metrics.
}
    \label{fig:2d_hist}
\end{figure*}

The 2D histogram plotting log probabilities for energy, time and energy \& time against LOR TOF estimated annihilation point's distance from point source center are for both isotopes for one of the configurations: 100 ps timing FWHM, 12\% energy resolution FWHM, 4 mm spatial FWHM is shown in Fig.~\ref{fig:2d_hist}. The high density within in lower right corner indicates most probable LORs are also located close to the point source.

\begin{figure*}[htbp]
    \centering
    \includegraphics[width=2 \columnwidth]{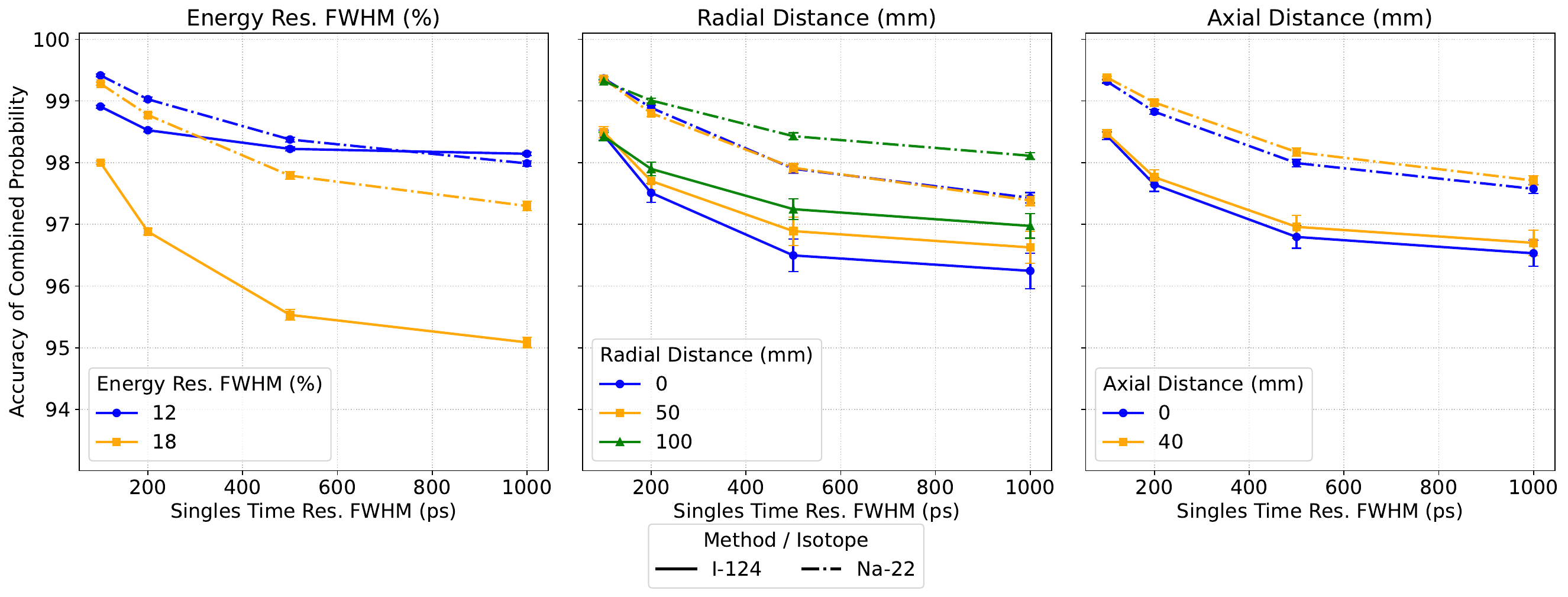}
    \caption{MLE Energy \& Time accuracy versus singles time resolution for two isotopes. The three panels isolate dependence on (left) energy resolution FWHM, (middle) radial source position, and (right) axial source position, with all curves shown for spatial FWHM = 4 mm. Each point is the mean accuracy across seeds and error bars denote SEM. Colors/markers identify the panel-specific varying factor, while line style distinguishes isotopes (I-124 and Na-22).}
    \label{fig:three_panel_point_source}
\end{figure*}

\begin{figure}[htbp]
    \centering
    \includegraphics[width=\linewidth]{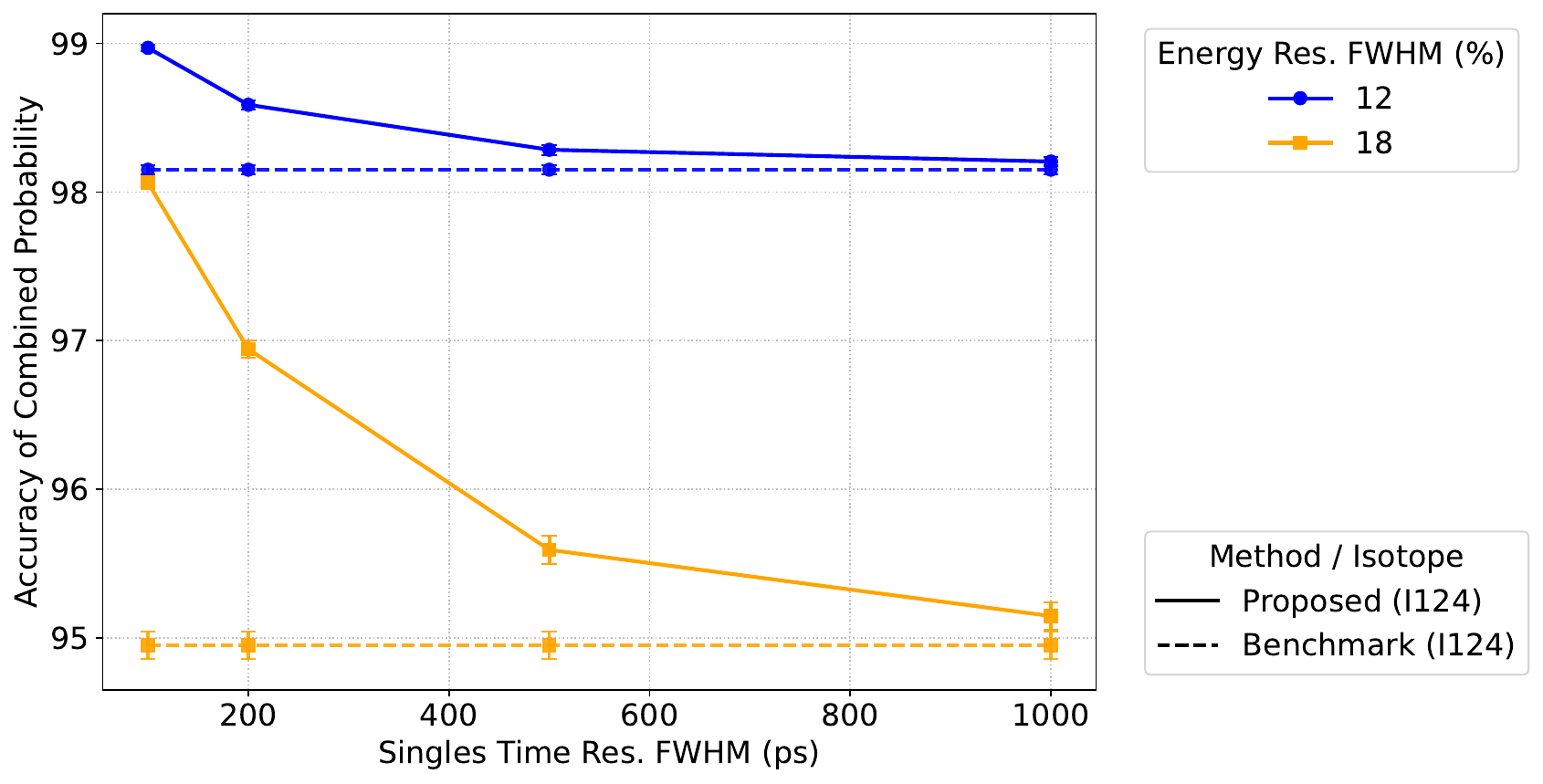}
    \caption{Mean $\pm$ standard deviation (error bars) of the MLE energy \& time accuracy for a $^{124}$I point source at various positions, evaluated across different energy and singles time resolutions. Benchmark method is shown with transparent lines for the two different energy resolutions.}
    \label{fig:i124_point_benchmark}
\end{figure}

The MLE energy \& time accuracy for both isotopes as a function of energy, radial and axial distance over singles time resolutions is shown in Fig.~\ref{fig:three_panel_point_source}. The algorithm yields above 96\% accuracy for all $^{22}$Na point source positions for all time and energy resolutions. The large role timing resolution plays in accuracy is evident by the sharp dip in accuracy as time resolution gets worse. The benchmark method for ${}^{22}$Na of selecting the prompt-gamma $\geq$ 650 and requiring the annihilation photons to be in 650-450, had nearly the same result of 100\% accuracy however we lose as much as 8\% of ``possible'' triple coincidences, as shown in Table \ref{tab:na22_possible_benchmark_counts}. For ${}^{124}$I along with the performance of the benchmark method (since for ${}^{124}$I case the benchmark conditions are more relaxed and benchmark can be applied to any triple coincidence) is shown in Fig.~\ref{fig:i124_point_benchmark}. The accuracy difference between 12 \% and 18 \% energy resolution is considerably larger than for $^{22}$Na point source (left) Fig.~\ref{fig:three_panel_point_source}. This is due to $^{124}$I's prompt-gamma energy of 602 keV being closer to the 511 keV annihilation energy than $^{22}$Na's prompt-gamma energy of 1275 keV resulting in energy resolution having a larger effect on accuracy. 

\begin{table}[htbp]
\centering
\footnotesize
\setlength{\tabcolsep}{3pt}
\begin{tabular}{rr rr SS}
\toprule
\multicolumn{2}{c}{Position (mm)} &
\multicolumn{2}{c}{Counts} &
\multicolumn{2}{c}{Triple Fraction} \\
\cmidrule(lr){1-2}
\cmidrule(lr){3-4}
\cmidrule(lr){5-6}
$x_c$ & $z_c$ &
Total & Possible &
\multicolumn{1}{c}{Possible (\%)} &
\multicolumn{1}{c}{Benchmark (\%)} \\
\midrule
0   & 0   & 1579 & 1471 & 93.16 & 85.78 \\
0   & 40  & 1728 & 1538 & 89.01 & 84.23 \\
50  & 0   & 1464 & 1331 & 90.92 & 84.42 \\
50  & 40  & 1981 & 1762 & 88.95 & 85.09 \\
100 & 0   & 2136 & 1880 & 88.02 & 86.99 \\
100 & 40  & 3147 & 2737 & 86.97 & 85.23 \\
\bottomrule
\end{tabular}
\caption{Benchmark counts and percentages for ${}^{22}$Na. 
``Possible (\%)'' denotes the fraction of triple coincidences for which at least one LOR lies within 10~mm of the source (non-blurred reference). 
``Benchmark (\%)'' denotes the fraction of triple events satisfying the benchmark condition: one photon above 650~keV and at least one possible LOR within the FOV.}
\label{tab:na22_possible_benchmark_counts}
\end{table}

\begin{table}[htbp]
\centering
\footnotesize
\setlength{\tabcolsep}{3pt}
\begin{tabular}{rr rr SS}
\toprule
\multicolumn{2}{c}{Position (mm)} &
\multicolumn{2}{c}{Counts} &
\multicolumn{2}{c}{Fraction of Total Triples} \\
\cmidrule(lr){1-2}
\cmidrule(lr){3-4}
\cmidrule(lr){5-6}
$x_c$ & $z_c$ &
Total & Possible &
\multicolumn{1}{c}{Possible (\%)} &
\multicolumn{1}{c}{Benchmark (\%)} \\
\midrule
0   & 0   & 4806 & 4142 & 86.18 & 96.57 \\
0   & 40  & 3269 & 2659 & 81.34 & 94.76 \\
50  & 0   & 4902 & 4037 & 82.35 & 94.68 \\
50  & 40  & 3742 & 3000 & 80.17 & 93.75 \\
100 & 0   & 6812 & 5318 & 78.07 & 88.97 \\
100 & 40  & 5896 & 4572 & 77.54 & 88.88 \\
\bottomrule
\end{tabular}
\caption{Benchmark counts and percentages for ${}^{124}$I. ``Possible (\%)'' denotes the fraction of triple coincidences for which at least one LOR lies within 10~mm of the source (non-blurred reference). ``Benchmark (\%)'' denotes the fraction of triple events for which at least one LOR lies within the FOV.}
\label{tab:i124_possible_benchmark_counts}
\end{table}

\begin{table}[htbp]
\centering
\setlength{\tabcolsep}{5pt}
\caption{Accuracy (\%) for energy-only and time-only probability methods (mean $\pm$ std).}
\begin{tabular}{c|cc|cccc}
\toprule
 & \multicolumn{2}{c|}{\shortstack{Energy Res.\\FWHM (\%)}} 
 & \multicolumn{4}{c}{\shortstack{Singles Time Res.(ps)}} \\
\midrule
 & 12 & 18 
 & 100 & 200 & 500 & 1000 \\
\midrule
${}^{22}$Na
& \shortstack{97.74 \\ $\pm$0.39} & \shortstack{96.97 \\ $\pm$0.62}
& \shortstack{97.15 \\ $\pm$0.52} & \shortstack{96.67 \\ $\pm$0.53}
& \shortstack{94.29 \\ $\pm$0.79} & \shortstack{89.92 \\ $\pm$1.41} \\

${}^{124}$I
& \shortstack{98.11 \\ $\pm$0.23} & \shortstack{94.90 \\ $\pm$0.64}
& \shortstack{93.54 \\ $\pm$0.63} & \shortstack{93.14 \\ $\pm$0.66}
& \shortstack{90.94 \\ $\pm$1.00} & \shortstack{86.85 \\ $\pm$1.73} \\
\bottomrule
\end{tabular}
\label{tab:energy_time_possible_only}
\end{table}

\subsection{Heterogeneity Phantom}
\begin{figure*}[htbp]
    \centering
    \includegraphics[width=\linewidth]{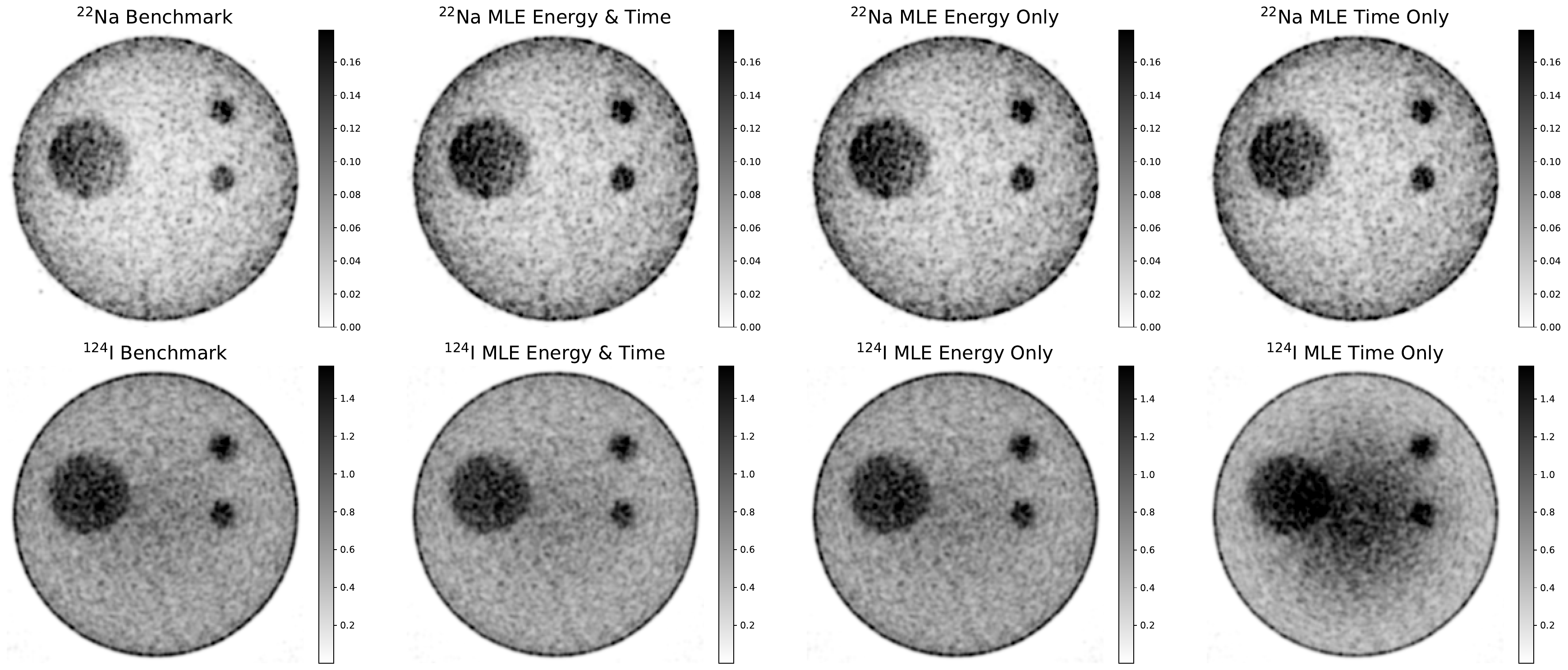}
    \caption{Axial slices (averaged over the central 5 mm) of triple-coincidence images of the heterogeneity phantom (Fig.~\ref{fig:6_sphere_layout}), reconstructed using benchmark and MLE LOR assignment methods: MLE energy \& time (combined), MLE energy-only, and MLE time-only. The top row corresponds to ${}^{22}$Na and the bottom row to ${}^{124}$I.}
    \label{fig:hetro_phantom_comparison}
\end{figure*}

Triple coincidence images of the heterogeneity phantom, illustrated in Fig.~\ref{fig:6_sphere_layout}, are shown in Fig.~\ref{fig:hetro_phantom_comparison} for both benchmark and MLE LOR reconstructions. The MLE reconstructions include methods incorporating both time and energy information, energy only, and time only. Quantitative evaluation of the reconstructions is provided in Table~\ref{tab:quant_comparison}, which reports CRC values for ROIs containing the $\beta^+\!\!-\!\!\gamma$ emitter and XR values for ROIs containing only the $\beta^+$ emitter. 

Bootstrap analysis was used to assess statistical significance of differences between reconstruction methods for both isotopes. For ${}^{22}$Na, most differences between the benchmark and MLE-based methods were not statistically significant, as the 95\% confidence intervals included zero, indicating comparable performance. The only exception was CRC1 (6 cm), where both MLE energy-only and MLE time-only showed a small but consistent degradation relative to the benchmark; no significant differences were observed for XR metrics. In contrast, for ${}^{124}$I, performance differences were more pronounced across methods. MLE energy \& time and MLE energy-only were statistically indistinguishable from the benchmark for all CRC and XR metrics,  and MLE time-only exhibited significant degradation in all CRC and XR. Overall, these results suggest that incorporating energy information—either alone or combined with time—yields performance comparable to the benchmark, while reliance on timing information alone leads to statistically significant degradation, particularly for ${}^{124}$I.

\begin{table*}[htbp]
\centering
\setlength{\tabcolsep}{4pt}
\renewcommand{\arraystretch}{1.1}
\begin{tabular}{lcccccc}
\toprule
\textbf{Method} & \textbf{Isotope} & \textbf{CRC 1} & \textbf{CRC 3} & \textbf{CRC 4} & \textbf{XR 2} & \textbf{XR 5} \\
 &  & \textbf{(6 cm)} & \textbf{(2 cm)} & \textbf{(2 cm)} & \textbf{(6 cm)} & \textbf{(2 cm)} \\
\midrule
Benchmark & ${}^{22}$Na & $\mathbf{0.619 \pm 0.001}$ & $\mathbf{0.733 \pm 0.004}$ & $\mathbf{0.560 \pm 0.003}$ & $0.071 \pm 0.002$ & $0.340 \pm 0.008$ \\
MLE Energy \& Time & ${}^{22}$Na & $0.598 \pm 0.001$ & $0.673 \pm 0.003$ & $0.530 \pm 0.003$ & $\mathbf{0.064 \pm 0.002}$ & $0.251 \pm 0.007$ \\
MLE Energy Only & ${}^{22}$Na & $0.586 \pm 0.001$ & $0.654 \pm 0.003$ & $0.521 \pm 0.003$ & $0.077 \pm 0.001$ & $0.283 \pm 0.007$ \\
MLE Time & ${}^{22}$Na & $0.592 \pm 0.001$ & $0.659 \pm 0.003$ & $0.524 \pm 0.003$ & $0.072 \pm 0.002$ & $\mathbf{0.236 \pm 0.007}$ \\
\midrule
Benchmark & ${}^{124}$I & $0.270 \pm 0.000$ & $0.228 \pm 0.001$ & $0.190 \pm 0.001$ & $0.023 \pm 0.001$ & $\mathbf{-0.078 \pm 0.004}$ \\
MLE Energy \& Time & ${}^{124}$I & $\mathbf{0.273 \pm 0.001}$ & $\mathbf{0.236 \pm 0.001}$ & $0.190 \pm 0.002$ & $\mathbf{-0.022 \pm 0.001}$ & $-0.128 \pm 0.004$ \\
MLE Energy Only & ${}^{124}$I & $0.270 \pm 0.001$ & $0.228 \pm 0.001$ & $\mathbf{0.191 \pm 0.001}$ & $0.025 \pm 0.001$ & $-0.080 \pm 0.004$ \\
MLE Time & ${}^{124}$I & $0.172 \pm 0.000$ & $0.101 \pm 0.001$ & $0.111 \pm 0.001$ & $-0.089 \pm 0.001$ & $-0.355 \pm 0.003$ \\
\bottomrule
\end{tabular}
\caption{Quantitative comparison of ${}^{22}$Na and ${}^{124}$I heterogeneity phantom reconstructions. CRC values are reported for ROIs containing $\beta^+\!\!-\!\!\gamma$ activity (ROIs 1, 3, and 4), and XR values are reported for $\beta^+$ ROIs (ROIs 2 and 5). Sphere diameters are shown in parentheses. Higher CRC indicates better recovery, while XR values closer to zero indicate better cross-talk suppression. \textbf{Bold values denote the best performance for each isotope}. Values represent mean ± standard error of the mean (SEM) computed from 100 bootstrap reconstructions generated by resampling the list-mode data with replacement.}
\label{tab:quant_comparison}
\end{table*}

\section{Discussion}
Overall, the proposed MLE LOR assignment framework achieves performance comparable to or exceeding the benchmark method across both isotopes. Energy information provides strong discrimination for LOR assignment, while the incorporation of timing information further improves performance under TOF conditions. Differences between isotopes highlight the impact of underlying emission physics on achievable accuracy.

In point-source simulations, energy-only reconstruction achieves accuracies exceeding 94\% for both isotopes (Table~\ref{tab:energy_time_possible_only}), with additional gains observed when incorporating timing information (Fig.~\ref{fig:three_panel_point_source}). Time-only reconstruction achieves high accuracy for ${}^{22}$Na under TOF conditions ($\leq 500$~ps), but performs worse for ${}^{124}$I. This difference is attributable to the higher mean positron energy and larger positron range of ${}^{124}$I (825.9~keV and 3.4~mm in water) compared to ${}^{22}$Na (220.3~keV and 0.53~mm) \cite{jodal_positron_2014}, which degrades spatial consistency and TOF-based discrimination.

The benchmark method exhibits different behavior across isotopes due to its selection criteria. For ${}^{22}$Na, the requirement of a high-energy photon ($\geq 650$~keV) leads to a loss of usable events (Table~\ref{tab:na22_possible_benchmark_counts}); when these conditions are satisfied, both benchmark and MLE methods achieve nearly perfect accuracy, but outside these conditions the MLE energy-only method remains correct in more than 75\% of cases. In contrast, for ${}^{124}$I the relaxed benchmark condition (requiring only one LOR within the FOV) avoids this sensitivity limitation, enabling direct comparison (Fig.~\ref{fig:i124_point_benchmark}) and demonstrating improved performance of the combined MLE energy-and-time approach over selecting the maximum-energy photon alone.

In the heterogeneity phantom (Fig.~\ref{fig:hetro_phantom_comparison}), the MLE and benchmark reconstructions for ${}^{22}$Na are visually and quantitatively similar, with no statistically significant differences in CRC or XR. However, the MLE methods produce higher event counts, which may be beneficial in low-sensitivity scenarios. For ${}^{124}$I, the benchmark, MLE energy \& time, and MLE energy-only methods show comparable performance, whereas the MLE time-only method results in degraded image quality and reduced quantitative accuracy. Although improvements in LOR assignment accuracy are observed in point-source studies for ${}^{124}$I (Fig.~\ref{fig:i124_point_benchmark}), these gains do not fully translate to the larger phantom. This discrepancy may be due to factors not accounted for in the current framework, including scatter and triple-coincidence sensitivity variations.

The point-source simulations were performed in a small (10~mm radius) water phantom and therefore do not capture patient-scale attenuation and Compton scatter. Under these conditions, low-energy events primarily arise from detector energy resolution and partial energy deposition rather than true scatter. In clinical imaging, increased scatter would broaden the detected energy spectrum and increase the likelihood of misclassifying scattered photons as annihilation or prompt-gamma candidates. This effect is expected to degrade the performance of the benchmark method, which relies on energy thresholding. While the MLE framework would also be affected through the energy probability term, the inclusion of timing and spatial consistency constraints may provide partial robustness by penalizing physically inconsistent events. Evaluation in larger anthropomorphic phantoms is required to fully characterize performance under realistic imaging conditions. For experimental implementation, additional detector-level corrections are required. In particular, time walk effects associated with high-energy prompt gamma must be addressed. Due to the energy dependence of detector timing response, prompt gamma events (e.g., 602–1275~keV) can introduce systematic timing offsets relative to 511~keV annihilation photons, which will result in inaccurate timestamps if uncorrected \cite{xie_methods_2020}. Accurate time walk correction is therefore necessary to fully realize the benefits of incorporating timing information in the proposed MLE framework.

The observed dependence on time and energy resolution suggests opportunities for further optimization of the MLE framework. The current implementation assigns equal weighting to energy and timing probabilities; however, system-specific weighting could improve performance and may be determined through parameter optimization (e.g., grid search). Additionally, thresholding on posterior probabilities could be used to balance accuracy and sensitivity by rejecting low-confidence LOR assignments.

The modular formulation of the MLE framework enables straightforward incorporation of more detailed physical models. For example, system-dependent TOF kernels and spatial resolution models can be incorporated to account for variations across detector positions. The framework can also be extended to include additional effects such as acollinearity, by relaxing the collinearity constraint between photon directions, and positron range, by introducing additional latent variables in the spatial probability model (Eq.~\ref{eq:pos_gaussian}). These extensions provide a pathway toward improved physical accuracy without fundamentally altering the reconstruction framework.

Finally, the performance observed for ${}^{22}$Na and ${}^{124}$I suggests that the proposed method is applicable to a broader class of positron–prompt-gamma emitters, including ${}^{44}$Sc, ${}^{72}$As, and ${}^{86}$Y. These isotopes have prompt-gamma energies within the range explored in this work (602–1275~keV) and generally exhibit shorter positron ranges than ${}^{124}$I, indicating that similar or improved performance may be achievable.

\section{Conclusion}
In this work, we explored a maximum-likelihood estimation (MLE) framework for assigning the correct LOR in triple coincidences of positron-prompt-gamma emitters. For both isotopes, while energy-only reconstruction achieves accuracies exceeding 94\%, incorporating time information further improves performance. Our MLE algorithm achieves $>$94\% accuracy for $^{22}$Na and $>$90\% for $^{124}$I across all tested energy, timing, and spatial resolutions and positions. In non-TOF systems, energy information provides substantial improvement over timing information alone, highlighting its importance for accurate LOR assignment and providing guidance on scanner capabilities required for reliable mPET imaging. Finally, we demonstrate a practical application of this algorithm for triple coincidences in mPET combining a $\beta^+$ emitter and a $\beta^+\!\!-\!\!\gamma$ emitter, showing the ability to separate overlapping radionuclide signals in a warm background with comparable performance with benchmark.

\appendix
\section{Time Probability Threshold} \label{sec:time_threshold}
While the energy probability may be high for random events, annihilation photons from ${}^{18}$F /$\beta^+$ can occur in coincidence with a prompt gamma (recorded as an ${}^{18}$F event in the triple-coincidence image), or two annihilation photons may originate from different sources (the conventional PET random events). However, because the time probability assumes that annihilation photons are collinear and that all three photons originate from the same decay location, it can be used to filter out events where the photons do not come from the same source.

Using the heterogeneity phantom, we analyzed a small subset of the data and recorded both the time probability and the source ID of the annihilation photons assigned to the LOR. For ${}^{22}$Na, 1\% of events were conventional PET randoms, 2\% were from ${}^{18}$F, and 97\% were true events. For ${}^{124}$I, 4\% were conventional PET randoms, 5\% were ${}^{18}$F annihilation photons, and 91\% were true events.

\begin{figure}[htbp]
    \centering
    \includegraphics[width=\linewidth]{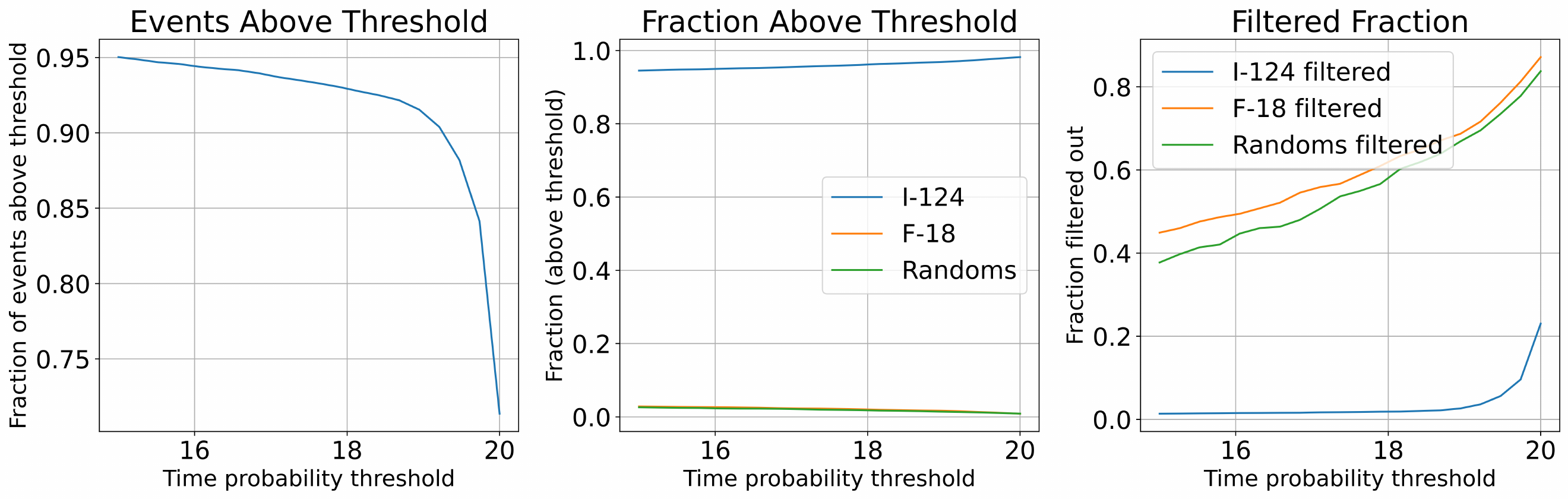}
    \caption{Left to right: percent of events above the log time probability threshold, proportion of events after threshold filtering, and percent of each event type filtered out above the threshold.}
    \label{fig:tt_graph}
\end{figure}

We observed that increasing the time threshold increases the relative proportion of true prompt-gamma positron-emitter events (Fig.~\ref{fig:tt_graph}). The trends were similar for both ${}^{22}$Na and ${}^{124}$I, and based on this analysis, we selected a time threshold of 19 for both isotopes to balance statistics with random event filtering.

\section*{Acknowledgments}
The authors declare no conflict of interest. Sarah Zou is the Mark and Mary Stevens Interdisciplinary Graduate Fellow at Stanford University. This work was supported in part by Dr. Ralph \& Marian Falk Medical Research Trust (SPO 268891), Stanford Bio-X Interdisciplinary Initiatives Seed Grants Program (IIP), and Mars Shot Award from the SNMMI.

\bibliographystyle{ieeetr}
\bibliography{Reference, references_zotero}

\end{document}